%% file: ms.tex
\documentclass{llncs}

\usepackage{lmodern}        
\usepackage[T1]{fontenc}    
\usepackage[utf8]{inputenc} 

\usepackage{amsmath}    
\usepackage{amssymb}    
\usepackage{mathtools}  
\usepackage{microtype}  
\usepackage[inline]{enumitem} 
\usepackage{multirow}   
\usepackage{booktabs}   
\usepackage{subcaption} 
\usepackage[usenames,dvipsnames,table]{xcolor} 
\usepackage{hyperref}  
\usepackage{xspace}

\author{Abraham Hinteregger} 

\def\TITLE{An Empirical Analysis of Monero Cross-Chain Traceability}
\hypersetup{
    pdfpagelayout   = TwoPageRight,           
    linkbordercolor = {Melon},                
    pdfauthor       = {Abraham Hinteregger},          
    pdftitle        = {\TITLE},         
    pdfkeywords     = {cryptocurrency, linkability, traceability, heuristics, empirical} 
}

\definecolor{vir0}{rgb}{0.267004, 0.004874, 0.329415}
\definecolor{vir1}{rgb}{0.282623, 0.140926, 0.457517}
\definecolor{vir2}{rgb}{0.253935, 0.265254, 0.529983}
\definecolor{vir3}{rgb}{0.206756, 0.371758, 0.553117}
\definecolor{vir4}{rgb}{0.163625, 0.471133, 0.558148}
\definecolor{vir5}{rgb}{0.127568, 0.566949, 0.550556}
\definecolor{vir6}{rgb}{0.134692, 0.658636, 0.517649}
\definecolor{vir7}{rgb}{0.266941, 0.748751, 0.440573}
\definecolor{vir8}{rgb}{0.477504, 0.821444, 0.318195}
\definecolor{vir9}{rgb}{0.741388, 0.873449, 0.149561}
\definecolor{vir10}{rgb}{0.993248, 0.906157, 0.143936}

\input{./data/numberMacros.tex}

\begin{document}
\title{\TITLE}
\author{{Abraham Hinteregger}\inst{1,2} \and {Bernhard Haslhofer}\inst{1}}
\institute{{Austrian Institute of Technology} \and {Vienna University of Technology}}
\maketitle

\input{./sections/Paper.tex}
\end{document}

%% file: data/numberMacros.tex
\def\XMRDATEF{2014-04-18\xspace}
\def\XMODATEF{2018-04-06\xspace}
\def\XMVDATEF{2018-05-03\xspace}
\def\XMRDATEL{2018-08-31\xspace}
\def\XMODATEL{2018-08-31\xspace}
\def\XMVDATEL{2018-08-31\xspace}

\def\XMRFIRST{1\xspace}
\def\XMOFIRST{1\,546\,600\xspace}
\def\XMVFIRST{1\,564\,966\xspace}
\def\XMRHEIGHT{1\,651\,346\xspace}

\def\XMOHEIGHT{1\,651\,728\xspace}
\def\XMVHEIGHT{1\,647\,778\xspace}

\def\XMRTX{4\,955\,908\xspace}

\def\XMOTX{146\,475\xspace}
\def\XMVTX{146\,215\xspace}
\def\XMRCB{1\,651\,347\xspace}

\def\XMOCB{105\,729\xspace}
\def\XMVCB{82\,814\xspace}
\def\XMRTXOUT{28\,878\,846\xspace}

\def\XMOTXOUT{198\,618\xspace}
\def\XMVTXOUT{450\,773\xspace}
\def\XMRR{24\,760\,168\xspace}

\def\XMOR{244\,965\xspace}
\def\XMVR{212\,919\xspace}
\def\XMRRNT{12\,538\,632\xspace}

\def\XMORNT{241\,464\xspace}
\def\XMVRNT{212\,919\xspace}
\def\XMRRNTL{4\,212\,422\xspace}

\def\XMORNTL{50\,861\xspace}
\def\XMVRNTL{7\,671\xspace}
\def\XMRRM{70\,767\,723\xspace}

\def\XMORM{1\,243\,479\xspace}
\def\XMVRM{1\,701\,036\xspace}
\def\XMRIM{16\,270\,257\xspace}

\def\XMOIM{230\,128\xspace}
\def\XMVIM{49\,035\xspace}
\def\XMRIR{16\,433\,958\xspace}

\def\XMOIR{54\,362\xspace}
\def\XMVIR{7\,671\xspace}

\def\recenta{2018-04-01\xspace}
\def\recentb{2018-08-31\xspace}
\def\recentnofork{25\,256\xspace}
\def\recentfork{73\,321\xspace}
\def\recentmixinnofork{203\,251\xspace}
\def\recentmixinfork{544\,131\xspace}
\def\recenttotal{617\,452\xspace}
\def\recentrm{11\,826\,525\xspace}
\def\recenttotalnf{228\,507\xspace}
\def\recentrings{1\,565\,858\xspace}
\def\recenttxnoncb{685\,608\xspace}



%% file: sections/Paper.tex

\begin{abstract}
	Monero is a privacy-centric cryptocurrency that makes payments untraceable by adding decoys to every real input spent in a transaction. Two studies from 2017 found methods to distinguish decoys from real inputs, which enabled traceability for a majority of transactions.
	Since then, a number protocol changes have been introduced, but their effectiveness has not yet been reassessed. Furthermore, little is known about traceability of Monero transactions across hard fork chains.
	We formalize a new method for tracing Monero transactions, which is based on analyzing currency hard forks. We use that method to perform a (passive) traceability analysis on data from the Monero, MoneroV and Monero Original blockchains and find that only a small amount of inputs are traceable.
	We then use the results to estimate the effectiveness of known heuristics for recent transactions and find that they do not significantly outperform random guessing.
	Our findings suggest that Monero is currently mostly immune to known passive attack vectors and resistant to tracking and tracing methods applied to other cryptocurrencies.
\end{abstract}

\section{Introduction}

Monero is a privacy-enhancing cryptocurrency that exceeds others (Zcash, Dash) in terms of market capitalization and promises privacy and anonymity through \emph{unlinkable} and \emph{untraceable} transactions.
It thereby addresses a central shortcoming of well-established currencies such as Bitcoin, which cannot offer a meaningful level of anonymity because transactions sent to addresses are linkable and payments among pseudonymous addresses are traceable. There are now a number of commercial (e.g., Chainalysis) and non-commercial tools~\cite{haslhofer_o_2016,kalodner_blocksci:_2017} that implement well-known analytics techniques (c.f.,~\cite{meiklejohn_fistful_2013}) and provide cryptocurrency analytics features, including tracking and tracing of payments made in cryptocurrencies.

Technically, Monero is based on the CryptoNote protocol and aims to address Bitcoin's privacy issues using three central methods: \emph{Stealth addresses}, which are one-time keys that are generated from the recipient's address and a random value, should prevent the identification of transactions sent to a given address and provide \emph{unlinkability}.
The use of \emph{Ring Signatures} in Monero transactions, which mixes an output that is spent (real input) with other decoy outputs (mixin input), obscures the path of a given coin and provide \emph{untraceability} of payments.
Finally, \emph{Confidential Transactions} hide the value of non-mining transactions and should prevent tracing by value and guessing of change addresses, which are used to send excess input funds back to the issuer of the transaction, based on values.

Nevertheless, in 2017, two concurrent studies~\cite{kumar_traceability_2017,moser_empirical_2018} have shown that \emph{untraceability} can be compromised by applying heuristics that can identify mixins.
They were able to trace the majority of transactions up to the introduction of RingCTs (confidential transactions) in Jan. 2017.
In the following releases (Sep. 2017 and Mar. 2018), additional improvements such as a higher mandatory minimal ringsize and an improved sampling technique for decoys were rolled out.
The 2017 studies already showed that the share of traceable transactions plummeted after the introduction of RingCTs, but a more up-to-date picture on the effectiveness of those improvements is still missing.

Furthermore, another traceability method that exploits information leaked by currency hard forks (a split of the currency with a shared history; unspent pre-fork TXOs can be spent on both post-fork branches) has been discussed in the community for some time. The general idea of the attack vector is to exploit differences between rings spending the same output on the two post-fork branches. However, we are not aware of a formal description, nor of an evaluation of its effectiveness.

The contributions of this work are twofold: first, in Section~\ref{sec:cca}, we propose and formalize a new Monero cross-chain analysis method, which exploits information leaked by currency hard forks.
Second, in Section~\ref{sec:results}, we empirically analyze Monero cross-chain traceability by combining known heuristics with our new method.
Our analysis, which considers Monero blocks 0 to \XMRHEIGHT (\XMRDATEL), also provides an up-to-date assessment on the effectiveness of the previously mentioned countermeasures by evaluating the accuracy of known heuristics on recent transactions.

Our findings suggest that Monero is currently mostly immune to known passive attack vectors and resistant to established tracking and tracing methods applied to other cryptocurrencies. They also confirm that the changes to the protocol, which were introduced as countermeasures to the heuristics proposed by Kumar et al.~\cite{kumar_traceability_2017} and Möser et al.~\cite{moser_empirical_2018}, were effective. Consequently, this implies that currently available cryptocurrency analytics tools that work for Bitcoin and its derivatives cannot be applied for tracking and tracing of Monero payments.

All the analysis done in this work can be reproduced by using the MONitERO toolchain, which can be found on GitHub\footnote{\url{https://github.com/oerpli/MONitERO}}

\section{Known Monero Traceability Methods}

Currently we are not aware of any known method to compromise confidential transactions and stealth addresses.
Untraceability has been successfully diminished by Kumar et al. and Möser et al.  with the following approaches:

\paragraph{Zero Mixin Removal (ZMR)}\cite{kumar_traceability_2017,moser_empirical_2018}
As each ring has exactly one real member, all rings with only one (non-mixin) member can be traced, just like in Bitcoin.
As the referenced outputs can only be spent once, occurrences of these outputs in other rings can be marked as \emph{mixins}.
Repeated applications of this rule is called \emph{Zero Mixin Removal (ZMR)}.
If the average ringsize is small enough, repeated applications of this rule can result in a chain reaction.
To prevent this, the mandatory minimum ringsize has been increased several times (0$\rightarrow$3 in 2016, 3$\rightarrow$5 in 2017 and 5$\rightarrow$7 in 2018).
In October 2018 (after the cutoff of our dataset) the ringsize has been increased (from 7$\to$11) and removed as parameter, i.e. all transactions issued since v0.13 have a fixed ringsize of 11.
This removes a possible attack vector, based on the assumption that transaction with certain nonstandard ringsizes are issued by the same (set of) users.
In our analysis we did not find a method to exploit this.
\paragraph{Intersection Removal (IR)}\cite{moser_empirical_2018}
This heuristic is a generalization of ZMR: If $N$ rings contain the same $N$ (non-mixin) members, it is (usually) impossible to determine, which output has been spent in which ring, but as all of them are spent in one of the rings, we mark them as \emph{spent} and other references to these outputs as mixin.
If $N=1$, this method is identical to ZMR.
This can be generalized even further: Instead of $N$ identical rings with $N$ members we look for sets $S$ of rings (where each ring is a set of transaction outputs, abbreviated as TXOs) with the property:	$|S| = \left|\bigcup_{r \in S} r\right|$ (if there are $n$ sets in $S$, the union of those sets contain $n$ elements).
This maps to the matching problem for bipartite graphs $G=(V_1,V_2,E)$, where the property $|S| \leq \left|N(S)\right|$ (where $N(x)$ is the set of neighbors of $x$) holds $\forall S: S\subseteq V_1$ iff there is a perfect matching.

\paragraph{Guess Newest Heuristic}\cite{kumar_traceability_2017,moser_empirical_2018}
While the outputs spent in a transaction are (mostly) fixed, the choice of decoys is somewhat arbitrary.
Most Monero TXOs are spent a few days after they've been received (resulting in an age distribution of the real inputs that is heavily left-skewed).
The mixins on the other hand were initially (until v0.9 in 2016) sampled uniformly from all eligible existing outputs.
Starting from January 2016, a triangular distribution was used (according to Möser et al. \cite{moser_empirical_2018}, which still wasn't sufficiently left-skewed and did not match real spending behavior), from December 2016 on, $\approx$25\% of the inputs where sampled from the \emph{recent zone} (outputs less than 5 days old).
In September 2017, two changes have been made to the sampling from the recent zone.
The recent zone has been reduced to outputs less than 1.8 days old and the number of decoys sampled from the recent zone has been increased to 50\%\footnote{\url{https://github.com/monero-project/monero/pull/1996/files}}.
In October 2018 (v0.13) the sampling method has been changed to a gamma distribution.
The two publications from April 2017 (shortly after the introduction of the recent zone sampling), proposed a simple yet highly effective heuristic, which guessed that the real input is the most recent one.

\paragraph{Output Merging Heuristic}\cite{kumar_traceability_2017}
If multiple inputs of a transaction reference distinct outputs from the same transaction, this heuristic assumes that these outputs are the real inputs.
Before the introduction of confidential transactions this arose naturally, as outputs where split up into denominations (e.g. an output of 8XMR would have been split up into three outputs with denominations 1,2 and 5 XMR).
If the recipient wants to spend his funds, he would have to merge these outputs.
While this heuristic also works for confidential transactions, these TXs tend to have fewer outputs (mostly 2 outputs, one of which is the change output), which results in fewer ``merging-transactions''.

\section{Hard Forks \& Cross Chain Analysis}\label{sec:cca}
\vspace*{-30pt}
\begin{figure}\centering
	\includegraphics{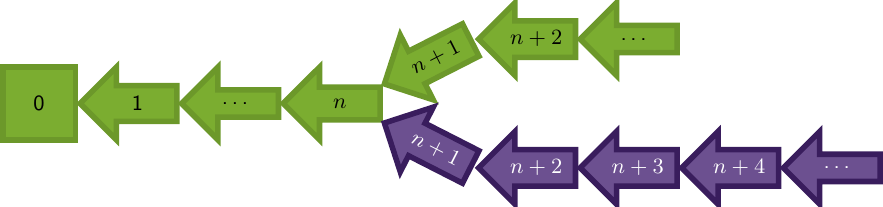}
	\caption{Illustration of a currency hard fork. The blocks between the genesis block (0) and the fork height (n) are shared. Unspent outputs from pre-fork TXs can be spent on both branches.}\label{fig:hardfork}
\end{figure}

\begin{figure}\centering
	\includegraphics{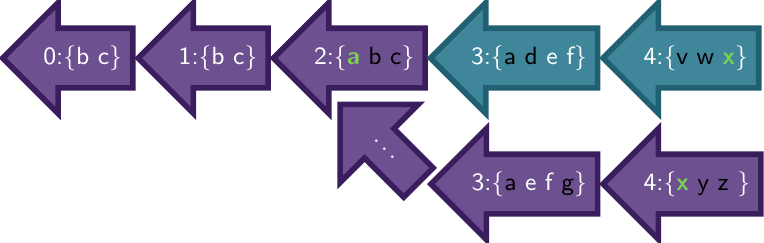}

\caption{An illustration of the Monero blockchain%
	(\protect\includegraphics{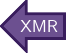}) and a fork of it (\protect\includegraphics{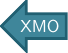}), two blocks before and the first two blocks after a fork.
	Each block contains one ring (format: "$\langle$KeyImg (0-9)$\rangle$\{$\langle$ring members (a-z)$\rangle$\}").
	The first two rings (0,1) have the same two members, i.e. intersection removal can be applied to mark these inputs ($b,c$) as mixin (\textcolor{black}{black}) in ring 2, leaving only input $a$, which is therefore the real (\textcolor{vir8}{\textbf{green}}) input.
	From the two rings with KeyImg 3, input $a$ can be therefore removed as it is spent.
	Additionally, $d$ and $g$ can also be ruled out as they are not part of the intersection $\{a,d,e,f\}\!\cap\!\{a,e,f,g\}$.
	The intersection of the two rings with KeyImg 4 consists of only one element, $x$, which must therefore be the real input.}
	\label{fig:methods}
\end{figure}

Like software projects, cryptocurrencies and their blockchains can be forked, resulting in two currencies with a partly shared transaction history.
There are different forking mechanisms, though for this work only hard forks (see Fig.~\ref{fig:hardfork}) are relevant.
The important aspect for our method is that unspent pre-fork TXOs can be spent on both branches.
To prevent double spending, each (input) ring is published with a \emph{key image}, which is uniquely determined by the spent output (the ring signature is used to verify that this is the case).
If a coin is spent in multiple (one per branch) rings after a fork, the real input has to be in the intersection of the rings and the remaining members, i.e. the pairwise symmetric differences, can be marked as mixins, as illustrated in Fig.~\ref{fig:methods}.

\section{Results}\label{sec:results}
\begin{figure}\centering
	\includegraphics{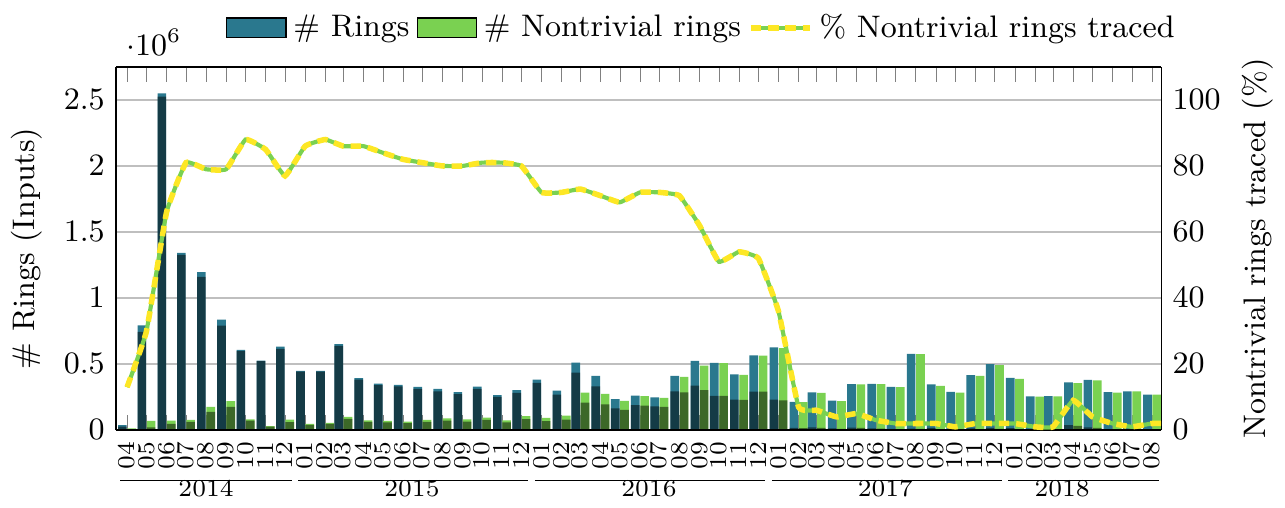}
	\caption[Monthly tracing statistics]{Bar chart of monthly number of nontrivial rings (NTR, $>1$ member). Shaded bars represent traced rings. Traceability plummets after introduction of RingCT, small peak after hard forks in Spring 2018 due to cross-chain analysis.}\label{fig:traced}
	\includegraphics{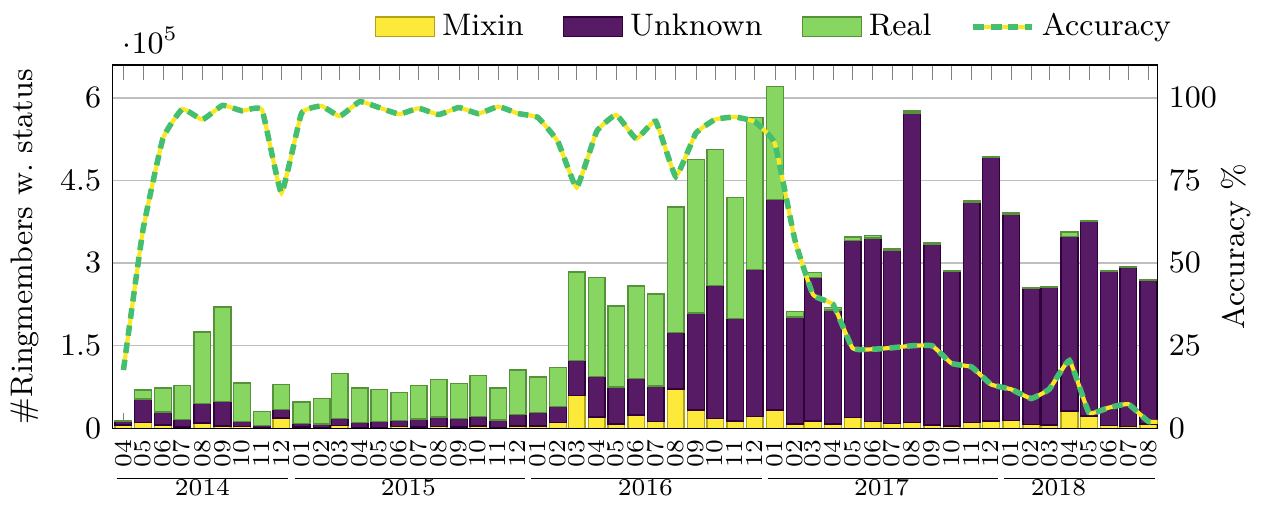}
	\caption[Guess Newest Heuristic: Performance over time]{\textbf{Performance of GNH over time:}
	After January 2017 the number of identified mixins and real inputs plummet and the accuracy is estimated based on a small sample. (Estimated) accuracy plummets for recent transactions.}\label{fig:gnh_accuracy}
	\includegraphics{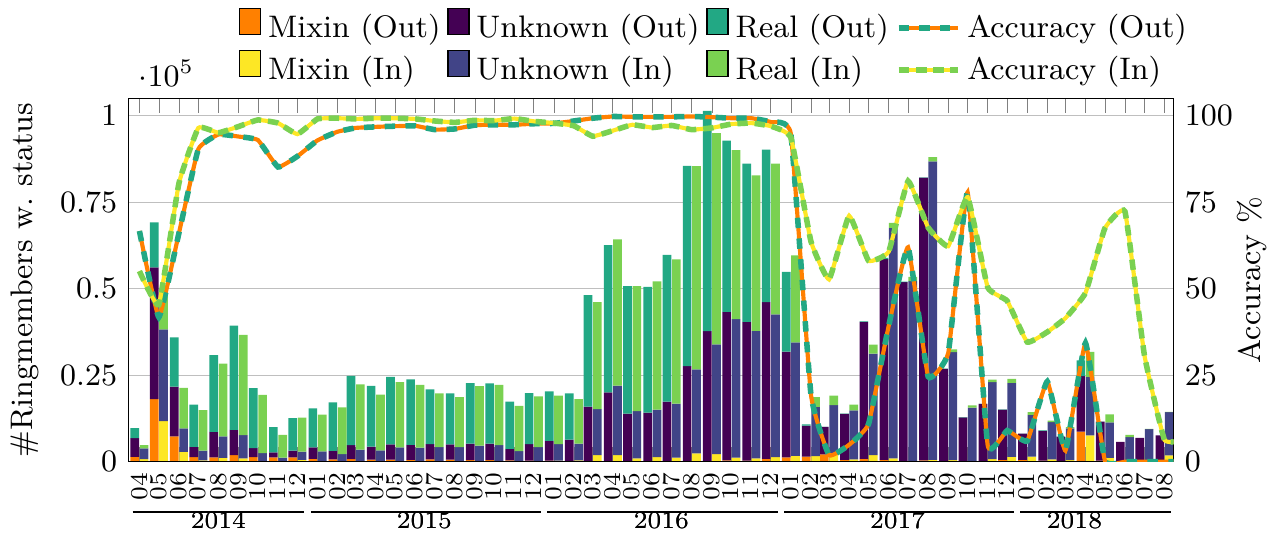}
	\caption[Output Merging Heuristic: Performance over time]{\textbf{Performance of OMH over time:} Outputs are created at time \emph{out} and spent at time \emph{in}. Left bar for each month uses \emph{out}-time for aggregation, right bar uses \emph{in}.}%
	\label{fig:merge_stats}
\end{figure}

We used the methods from Kumar et al. and Möser et al. that do not produce false positives (Zero Mixin Removal \& Intersection Removal) as well as our new method (Cross Chain Analysis) to deduce mixins and real inputs for Monero transactions.
We focus on \emph{nontrivial rings}, i.e. rings that have at least one mixin (ringsize > 1).
If some of the ring members are identified as mixin, we refer to the remaining number of possible real inputs as \emph{effective ringsize}.
A ring with an effective ringsize of 1 is \emph{traced}.
Statistics from our dataset can be found in Table~\ref{tab:dataTable}. 

Overall, were able to trace \XMRRNTL nontrivial rings.
As can be seen in Fig.~\ref{fig:traced}, in the first years of Monero's existence most nontrivial rings where traceable (as most mixins were spent in trivial rings).
To prevent this, mandatory minimum ringsizes have been introduced, though the sampling of older, provably spent outputs as mixins remained a problem.
Starting from 2017, the introduction of RingCT mostly eliminates this threat, as RingCT transactions only sample outputs from other RingCT transactions, all of which were issued after the introduction of mandatory minimum ringsizes.

In the weeks following the MoneroV and Monero Original hard forks, the fraction of traceable rings increases.
This is due to our newly proposed method, which allows us to identify
the real spent output of \recentfork{} (improved from \recentnofork{}) out of \recentrings{} transaction inputs in the \recenttxnoncb{} (non-coinbase) transactions that have been issued between \recenta{} and \recentb{}.
The number of identified mixins in this time span has also more than doubled, from \recentmixinnofork{} to \recentmixinfork{}.
Taken together, the status (real or mixin) of \recenttotal{} out of \recentrm ring members in this time frame has been identified, which amounts to 5.22\% (compared to \recenttotalnf{} and 1.93\% without cross-chain analysis).
Results from our traceability analysis can be found in Table~\ref{tab:resultTable}.

Using the results from our traceability analysis, we also investigated the accuracy of the guess newest heuristic (GNH) and the output merging heuristic (OMH) for recent transactions (see Table \ref{tab:resultTableRecent}).
We find that the performance of the GNH (see Fig.~\ref{fig:gnh_accuracy}) is not better than random guessing for recent transactions.
For the OMH (see Fig.~\ref{fig:merge_stats}) we used two different methods to aggregate the data by months, once considering the time where the outputs were created (``Out'') and when they where spent (``In'').
Overall, the number of true and false positives is identical, though the distributions over time differ somewhat.
A problem of the OMH is the fact that RingCT transactions have fewer inputs and outputs, resulting in less transactions that merge outputs from previous transactions and thus less possible applications of the OMH.

\begin{table}[tbp]\centering 
	\caption[Dataset statistics]{\textbf{Dataset statistics:} As the Monero (XMR), MoneroV (XMV) and Monero Original (XMO) blockchains share some parts, the values from the two forks (XMV \& XMO) only refer to data unique to their blockchain. ``Last block'' refers to the last block used for the analysis in this work.}
	\label{tab:dataTable}
	\begin{tabular}{@{}lrrr@{}}\toprule
	&\multicolumn{1}{c}{\textbf{XMR}}&\multicolumn{1}{c}{\textbf{XMO}}&\multicolumn{1}{c}{\textbf{XMV}} \\\midrule
	First TX date&\XMRDATEF&\XMODATEF&\XMVDATEF\\
	Last TX date&\XMRDATEL&\XMODATEL&\XMVDATEL\\
	First block&\XMRFIRST&\XMOFIRST&\XMVFIRST\\
	Last block&\XMRHEIGHT&\XMOHEIGHT&\XMVHEIGHT\\\midrule
	\# Transactions&\XMRTX&\XMOTX&\XMVTX\\
	\# Coinbase TXs&\XMRCB&\XMOCB&\XMVCB\\ \midrule
	\# TX outputs&\XMRTXOUT&\XMOTXOUT&\XMVTXOUT\\
	\# Rings (TX inputs)&\XMRR&\XMOR&\XMVR\\
	\# Nontrivial rings&\XMRRNT&\XMORNT&\XMVRNT\\
	\# Ring members&\XMRRM&\XMORM&\XMVRM\\
	\bottomrule
\end{tabular}
	\caption[Traceability results]{\textbf{Traceability results:} Overall results for Monero and its two forks as well as results for recent Monero TXs. Ringmembers are \emph{spent} if they have been found in an intersection set (i.e. spent but it is not known in which TX)}
	\label{tab:resultTable}
	\begin{tabular}{@{}lrrr@{}}\toprule
	&\multicolumn{1}{c}{\textbf{XMR}}&\multicolumn{1}{c}{\textbf{XMO}}&\multicolumn{1}{c}{\textbf{XMV}} \\\midrule
	\# Nontrivial rings&\XMRRNT&\XMORNT&\XMVRNT\\
	\# Ring members&\XMRRM&\XMORM&\XMVRM\\\midrule
	\# Traced nontrivial rings&\XMRRNTL&\XMORNTL&\XMVRNTL\\
	\# Identified mixin ringmembers&\XMRIM&\XMOIM&\XMVIM\\
	\# Identified real ringmembers &\XMRIR&\XMOIR&\XMVIR\\
	\# Identified spent ringmembers& 13\,240 & 0 & 0 \\
	\bottomrule
\end{tabular}
	\caption[Traceability results for recent TXs]{\textbf{XMR Traceability results for recent TXs:} Subset of the XMR dataset from Table \ref{tab:resultTable} restricted to TXs between \recenta{} and \recentb{}.}
	\label{tab:resultTableRecent}
	\begin{tabular}{@{}lr@{}}\toprule
		&\multicolumn{1}{c}{\textbf{XMR}}\\\midrule
	\# Nontrivial rings & \recentrings\\
	\# Identified real rm. w/o new method & \recentnofork\\
	\# Identified real rm. with new method & \recentfork\\
	\# Identified mixin rm. w/o new method & \recentmixinnofork\\
	\# Identified mixin rm. with new method & \recentmixinfork\\
	\bottomrule
\end{tabular}
\end{table}

\section{Discussion}

Before mandatory minimum ringsizes were introduced, most rings were traceable.
With increasing mandatory minimum ringsizes (2016/09) the percentage of traceable NTR dropped and since the introduction of RingCT (2017) only a small fraction of all transactions can be traced with blockchain analysis techniques.
While our newly proposed method enables the tracing of some additional transaction inputs, the overall impact from this attack vector seems to be small so far.
This could change when a Monero fork with considerably higher traction is launched, which would presumably result in more redeemed outputs.
Using the traced rings we looked at the accuracy of the GNH \& OMH and found that their performance suffered from the recent changes to the transaction protocol.
Though it may be the case that our analysis underestimates the accuracy of the GNH, as most of the traced rings in recent months were traced with our new method, which identifies the real spent pre-fork output in post-fork transactions.
This skews the age of the spent outputs compared to regular usage.
Overall, the fraction of traceable rings remains low and we believe that unless additional attack vectors emerge, Monero remains resistant to analysis methods which have been applied to other cryptocurrencies.

\section*{Acknowledgments}

This work was partly funded by the European Commission through the project TITANIUM (Project ID: 740558) and by Austrian Research Promotion Agency (FFG) through the project VIRTCRIME (Project ID: 860672)

\bibliographystyle{plain}
\vspace*{-10pt}
\bibliography{crypto}